\documentclass[%
 reprint,
 amsmath,amssymb,
 aps,
]{revtex4-2}

\usepackage{graphicx}
\usepackage{dcolumn}
\usepackage{bm}
\usepackage{color}

\begin{document}

\preprint{APS/123-QED}

\title{Thermal transport characteristics of Fermi-Pasta-Ulam chains undergoing soft-sphere type collisions}

\author{Sankhadeep Bhattacharyya}
 \affiliation{Department of Mechanical Engineering, Indian Institute of Technology Kharagpur, West Bengal, India - 721302.}
\author{Puneet Kumar Patra}%
 \email{puneet.patra@civil.iitkgp.ac.in}
\affiliation{%
Department of Civil Engineering and Center for Theoretical Studies, Indian Institute of Technology Kharagpur, West Bengal, India - 721302}%

\date{\today}

\begin{abstract}
We show numerically that including soft-sphere type collisions in the celebrated Fermi-Pasta-Ulam ($FPU$) chain completely alters the thermal transport characteristics. The resulting $FPU^C$ chains, while being momentum preserving, satisfy the Fourier's law and do not show anomalous thermal transport behavior. Collisions play a significant role in reducing the boundary jumps typically observed in $FPU$ chains. The thermal conductivity of the $FPU^C$ chains is significantly smaller than the $FPU$ chains at low temperatures due to the fast redistribution of energy from the lowest mode of vibrations to the higher modes. At high temperatures, however, the $FPU^C$ chains have larger thermal conductivity than the $FPU$ chains due to the large contributions to the heat flux because of the large-magnitude short-ranged anharmonic collision force. 
\end{abstract}
\maketitle

One of the important questions concerning thermodynamics is the origins of Fourier's law:
\begin{equation}
J = -\kappa{\nabla T},
\label{eq:1}
\end{equation}
at microscopic scales. At macroscopic scales, the heat flux, $J$, is proportional to the temperature gradient, $\nabla T$, with the proportionality constant given by $\kappa$, the thermal conductivity which is a material property independent of the system dimensions. However, it is now known that at atomistic scales, certain low-dimensional systems exhibit anomalous thermal conduction wherein the Fourier's law is violated \cite{fujii2005measuring,balandin2008superior}. The origin of such behavior has typically been investigated using simple one-dimensional chains, which show a gamut of thermal transport characteristics. Consider a one-dimensional chain whose $i^{th}$ particle has a mass $m$ and momentum $p_i$. Let this particle interact with its nearest neighbors through a harmonic potential, $V_H$, and an anharmonic potential, $V_A$, both of which are distance,  $\Delta x$, dependent. Let the particle also be tethered to its initial position through an anharmonic tethering potential, $U(x_i)$. It is interesting to note that although, different one-dimensional chains can be obtained from this generalized situation, whose Hamiltonian is:
\begin{equation}
    \mathcal{H} = \sum\limits_{i=1}^{N} \left[\frac{p_i^2}{2m} + V_H(\Delta x_{i-1,i}) + V_A(\Delta x_{i-1,i}) + U(x_i) \right],
    \label{eq:2}
\end{equation}
their thermal transport characteristics are completely different. For example, a one-dimensional chain of $N$ harmonically coupled oscillators, obtained by substituting $V_A(\ldots) = U (\ldots) = 0$, shows ballistic thermal transport with $\kappa \sim N$ \cite{dhar2008heat}. When $V_A(\Delta x_{i-1,i}) = \frac{1}{4}c_1 \Delta x_{i-1,i}^4$ and $U(x_i) = 0$ one obtains a Fermi-Pasta-Ulam ($FPU$) chain that displays anomalous thermal transport characteristics with $\kappa \sim N^{\alpha}, \alpha > 0$ \cite{lepri1997heat}. With $V_A = 0$ and $U(x_i) = \frac{1}{4}c_2 x_i^4$, a $\Phi^4$ chain \cite{chen1996breather,aoki2000bulk,aoki2000non,bhattacharyya2020thermal} with normal thermal transport characteristics is obtained. Other simple one-dimensional chains like the  Frenkel-Kontorova (FK) chain \cite{hu1998heat,hu2005heat} may also be obtained by small modifications to the generalized Hamiltonian. 

Based on the ding-a-ling model proposed by Casati et. al \cite{casati1984one}, wherein every second particle is attached to its initial position with harmonic spring and the intermediate particle remains free, researchers initially thought that the presence of chaos is the key differentiator. Later, it was found that although necessary, the presence of chaos is not sufficient for ensuring normal thermal transport characteristics. This can be exemplified by the fact that both the $FPU$ and $\Phi^4$ chains are chaotic but have different characteristics. A comparison of the Hamiltonian for the two chains indicates that while the $FPU$ chain is momentum conserving, all chains with normal thermal transport characteristics (like the ding-a-ling, FK, and $\Phi^4$ chains) are momentum non-conserving. Researchers, therefore, attributed the anomalous thermal transport behavior to the momentum conserving nature of the chains. It was argued that in momentum conserving chains the energy transported by low-frequency long-wavelength vibration modes diffuse very slowly to other modes \cite{prosen2000momentum} ensuring long-time correlations in the chains. This energy transport mechanism gets disrupted in the presence of tethering potential. However, recently a handful of momentum conserving chains have been proposed that follow normal thermal transport characteristics \cite{savin2014thermal,giardina2000finite,gendelman2000normal,lee2010momentum,giardiana2005momentum}. Note that there are still confusions associated with the thermal transport characteristics of chains with asymmetric asymptotic free potentials, such as Lennard-Jones, where researchers have obtained contradicting results \cite{savin2014thermal,li20151d}.

Building onto the recent progress in developing one-dimensional momentum conserving chains with finite thermal conductivity, we propose a simple modification to the $FPU$ chain by incorporating soft-sphere collisions, and show numerically that the resulting chain, termed as the $FPU^C$ chain, follows the Fourier's law despite being momentum conserving. While doing so, we highlight several important changes that occur because of including collisions: the boundary thermal resistance decreases, the modal energy distribution is significantly altered and the contributions of collision forces towards total heat flux supersedes that due harmonic forces at high temperatures.

Consider an $FPU$ chain of $N$ particles with $m=1$. Let the equilibrium spacing between the neighboring particles be $l_{eq}=1$. With $x_i$ denoting the position of $i^{th}$ particle and stiffness equalling unity, the harmonic part of Hamiltonian becomes $V_H (\Delta x_{i-1,i}) = 0.5 \times (x_{i-1}-x_i-l_{eq})^2$. Assuming a small non-linearity, $c_1 =0.1$, the anharmonic part of the potential, $V_A(\Delta x_{i-1,i}) = 0.25 \times 0.1 \times \Delta x_{i-1,i}^4$. If the first and the last particle of the chain are connected to fixed particles, denoted by index 0 and $N+1$, the total Hamiltonian  may be rewritten as:
\begin{equation}
    \begin{array}{rcl}
         \mathcal{H}_\text{FPU} & = & \sum\limits_{i=1}^{N}p_i^2 + \sum\limits_{i=0}^{N}\left[ \dfrac{1}{2}(x_{i+1} - x_i - l_{eq})^2 \right. \\
         & & + \left. \dfrac{0.1}{4}(x_{i+1} - x_i - l_{eq})^4 \right]
    \end{array}
    \label{eq:3}
\end{equation}
This traditional $FPU$ Hamiltonian treats the particles as point masses, where two particles never undergo collisions with each other. If the assumption of point masses is relaxed, and a soft-sphere potential of the form: $V_A(\ldots) = \frac{a}{(x_i - x_{i-1})^6}$ is imposed, one gets a modified FPU chain with colliding and repelling particles. The Hamiltonian of this $FPU^C$ chain is given by:
\begin{equation}
    \begin{array}{rcl}
         \mathcal{H}_{\text{FPU}^C} & = & \mathcal{H}_\text{FPU} + \sum\limits_{i=0}^{N}\dfrac{a}{(x_{i+1} - x_{i})^6}
    \end{array}
    \label{eq:4}
\end{equation}
An order six potential has been used as it is able to prevent two particles from colliding while at the same time not taking too big a toll on the computation time. The constant $a$ is determined on the basis of the desired effective radius of the particles. Assuming, the effective radius of each particle as $r=0.025$, $a$ is chosen such that it approaches zero quickly beyond $2r = 0.05$, and be negligible at $2r$ vis-\'a-vis the harmonic potential. Considering these factors, the constant $a$ is taken to be $ 5 \times 10^{-10}$.

The $N$ particles of both $FPU$ and $FPU^C$ chains are initially arranged on a line at their equilibrium spacing and given an initial random velocity sampled from a uniform distribution of [-1,1]. A temperature gradient is introduced in the chains by keeping the first particle in contact with a reservoir of temperature $T_H$ and the last particle with a reservoir of temperature $T_C$, where $T_H > T_C$. The variation of these temperatures from the mean temperature $T_M =\frac{T_H + T_C}{2}$ is kept at 10\%. The simulations are carried out at different $T_M = 0.1, 0.5, 1,$ and 2 to understand the variation in the properties with increasing temperature. The reason for thermostatting only the first and the last particles is to identify the role of collisions in determining boundary thermal resistance. Both Nos\'e-Hoover thermostats \cite{hoover1985canonical} and Langevin thermostats have been used for temperature control. As both methods provide similar results, especially in large $\Delta T$ and $N$, we only show the results due to the Langevin thermostat here. The results for the Nos\'e-Hoover thermostat are shown in the supplemental materials. 


\begin{figure}[h]
    \centering
    \includegraphics[width=1.0\linewidth]{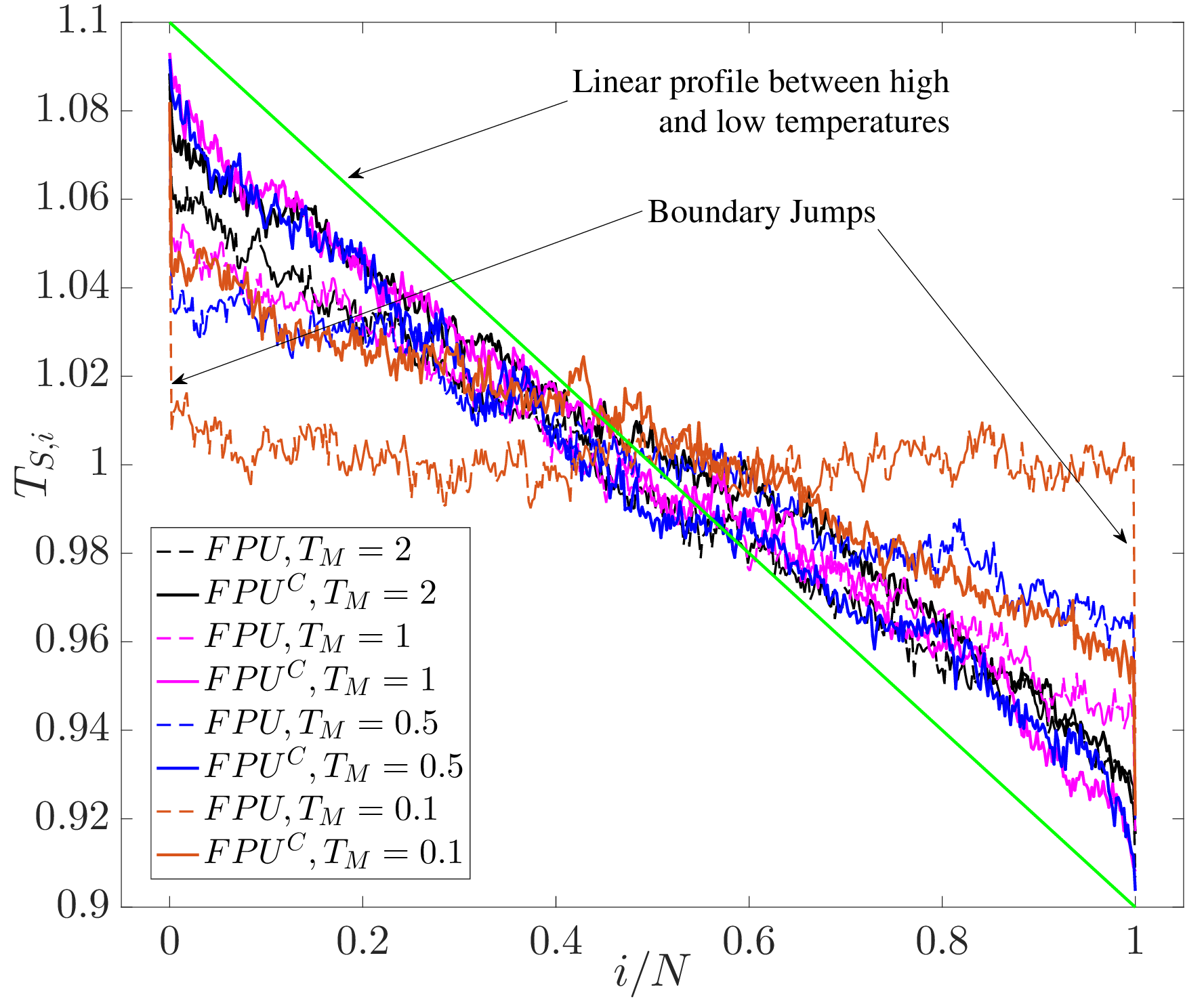}
    \caption{Normalized temperature profile across the $FPU$ and $FPU^C$ chains for different values of $T_M$. For any $i^{th}$ normalization is done through: $T_{S,i} = \langle T_i \rangle/T_M$.  Notice that the boundary jumps prevalent in the FPU chains are significantly smaller in $FPU^C$ chains. }
    \label{fig:TempProfile}
\end{figure}

Typically, under the assumption of local thermodynamic equilibrium, it is possible to define the different thermodynamic variables for every individual particle of a one-dimensional chain. Setting $k_B = 1$, one gets the instantaneous temperature of the $i^{th}$ particle as $T_i(t) = v_i^2(t)$, whose long time average $\langle T_i \rangle$ is used for plotting figure (\ref{fig:TempProfile}). The figure shows the normalized temperature profile of each particle of $T_{S,i} = \langle T_i \rangle/T_M$ for both $FPU$ and $FPU^C$ chains with $N=1024$. The deviation from a linear temperature profile (shown in green line) is smaller for $FPU^C$ chains than in $FPU$ chains. As has been reported previously \cite{aoki2001fermi}, boundary jumps in $FPU$ chains are more prominent at lower temperatures ($T_M = 0.1$  and 0.5) than at higher temperatures ($T_M = 2$). In $FPU^C$ chains, while we similarly observe boundary jumps at lower $T_M$, these jumps are significantly smaller than in $FPU$ chains. These boundary jumps represent the resistance offered to the heat transport due to the scattering of energy at the boundaries \cite{yilbas2017heat} and are not merely simulation effects. For $FPU^C$ chains, the boundary resistance is smaller because the collisions between the end thermostatted particles and the fixed boundary particles create an extra ``pinning'' effect \cite{candido2017eliminating}. Note that such reductions have been observed in colliding $\Phi^4$ chains as well \cite{bhattacharyya2020thermal}.  As the only difference between the two chains is the presence of collision potential, it is safe to conclude that the boundary resistance decreases substantially when collisions are modeled. 

\begin{figure}
    \centering
    \includegraphics[width=1.0\linewidth]{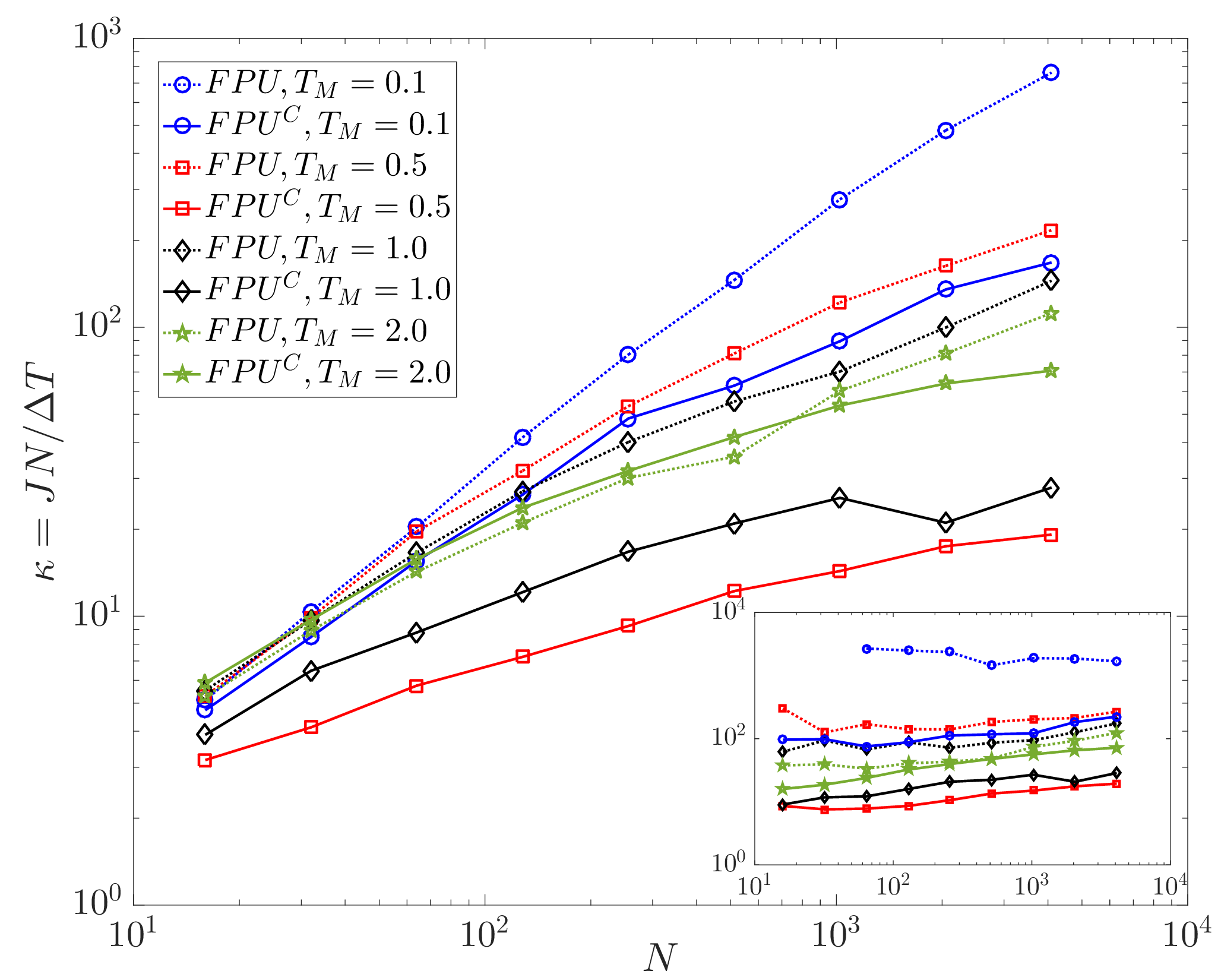}
    \caption{Thermal conductivity of $FPU$ and $FPU^C$ chains with $\Delta T$ equalling the imposed temperature difference (and actual temperature difference for the inset). $FPU^C$ chains obey Fourier's law of thermal conduction since its $\kappa$ saturates with increasing $N$. These scaling laws are not very prominent in the inset figure suggesting the important role played by the boundary jumps. In both cases, though, at low $T_M$, $\kappa$ for $FPU$ chains is larger than that in $FPU^C$ chains.}
    \label{fig:conductivity2}
\end{figure}

The local heat flux may be obtained from the time derivative of the local energy density. The instantaneous heat current at $i^{th}$ site can be written as $\dot{\epsilon}_i = \frac{\partial \epsilon_i}{\partial t} + \left[ j_{i-1,i}-j_{i,i+1} \right]$ where $j_{i,j}$ is the heat current between $i^{th}$ and $j^{th}$ particle. Under steady-state conditions, we get $\langle j_{i-1,i}\rangle$ = $\langle j_{i,i+1}\rangle$ since $\langle \dot{\epsilon} \rangle = \langle \frac{\partial \epsilon_i}{\partial t} \rangle =0$. Further, since $\left\langle \frac{dV(\Delta x_{i-1,i})}{dt} \right\rangle =0$, we get:
\begin{equation}
    \langle j_{i-1,i} \rangle = \left\langle \frac{1}{2}(v_i + v_{i-1})f_{i-1,i}\right\rangle = \langle v_if_{i-1,i} \rangle 
    \label{eq:hflux}
\end{equation}
The heat flux for the entire chain may be computed by: 
\begin{equation}
    J= \langle J \rangle = \frac{\langle \sum_{i=2}^{N} j_{i,i-1} \rangle}{N-1}
    \label{eq:hflux2}
\end{equation}
From $J$, the thermal conductivity can be computed as:
\begin{equation}
    \kappa = \frac{J(N-1)}{\Delta T}
    \label{eq:hflux3}
\end{equation}
In the presence of boundary jumps, a natural question arises about which value of $\Delta T$ should be used for computing $\kappa$. Taking $\Delta T$ as the difference between the imposed temperatures, one observes a thermal conductivity as shown in figure (\ref{fig:conductivity2}). Considering $\Delta T$ as the difference between the actual temperatures of the last but one particle from each end provides a heat flux as shown in the inset of figure (\ref{fig:conductivity2}). This approach effectively ignores the boundary jumps observed in figure (\ref{fig:TempProfile}). Note that the chain length has been reduced by two while computing $\kappa$. Several important deductions can be made from figure (\ref{fig:conductivity2}). The typical $FPU$ behavior -- $\kappa$ scales with increasing $N$, is observed when $\Delta T$ equals the imposed temperature difference, whereas, in $FPU^C$ chains, $\kappa$ appears to reach an asymptotic value with increasing $N$, suggesting that they obey the Fourier's law. Looking at the inset figure, it is evident that these scaling laws are not very obvious, especially for the $FPU$ chains. The boundary thermal resistance plays a major role here -- the top dotted line representing the $FPU$ chain at $T_M = 0.1$ has a slope almost equalling zero suggesting that $\kappa$ does not change with increasing $N$ which is contrary to what is known for the $FPU$ chains. In $FPU^C$ chains, the smaller values of boundary jumps ensure that $\kappa$ quickly achieves a scaling comparable with that when $\Delta T$ equals the imposed temperature difference. 

Interestingly, in both the cases, for $T_M < 2.0$, $\kappa$ of $FPU^C$ chains is smaller than that of $FPU$ chains. This has a well-grounded explanation from the kinetic theory-- $\kappa \sim \lambda c_s C_V$, where $\lambda$ is the mean free path, $c_s$ represents the speed of sound waves and $C_V$ denotes the heat capacity. Aoki and Kusnezov \cite{aoki2001fermi} have argued that in $FPU$ chains at low $T_M$, $\kappa \sim \lambda \sim 1/T_M$. In $FPU^C$ chains, because of the presence of soft-sphere collision terms, the mean free path is smaller, and hence, $\kappa$ is smaller as well. 

This can be understood better by looking at energy transport in Fourier space. Ignoring the anharmonic contributions arising due to the quartic interaction (and soft-sphere collision potential), one can write the dynamics of isolated $FPU$ and $FPU^C$ chains as $[\mathbf{M}]\mathbf{\ddot x} + [\mathbf{K}]\mathbf{x} = 0$, where, $[\mathbf{M}] = [\mathbf{I}]$ is the diagonal mass matrix, and the matrix $[\mathbf{K}]$ is the Hessian matrix with elements equalling $ K_{i,j} = \frac{\partial^2 V_H}{\partial x_i \partial x_j}$. In our case, $\mathbf{[K]}$ is symmetric with elements $K_{i,i} = 2$  and $K_{i,i+1} = K_{i-1,i} = -1$. Diagonalization of the mass-weighted stiffness matrix $[\mathbf{M}]^{-1}[\mathbf{K}]$ gives the normal-modal frequencies, $\omega_i^2, i \in [1,N]$, and the normal modes of vibration, $\vec{\xi}_i,i \in [1,N]$. The instantaneous modal displacements and momenta may be obtained by projecting the instantaneous displacement ($\mathbf{x}-\mathbf{x_0}$) and momentum ($\mathbf{p}$) of each particle on the eigenvectors ($\vec{\xi}_i$). The instantaneous energy of the $i^{th}$ normal mode may, therefore, be written as:
\begin{equation}
E_i(t) = \dfrac{1}{2}\omega_i^2 \delta_i^2(t) + \dfrac{1}{2} \dot{\delta}_i^2(t),
\label{eq:6}
\end{equation}
where, $\frac{1}{2}\omega_i^2 \delta_i^2(t)$ and $\frac{1}{2} \dot{\delta}_i^2(t)$, are the potential and kinetic energies of the $i^{th}$ mode, respectively. 
\begin{figure}
    \centering
    \includegraphics[width=\linewidth]{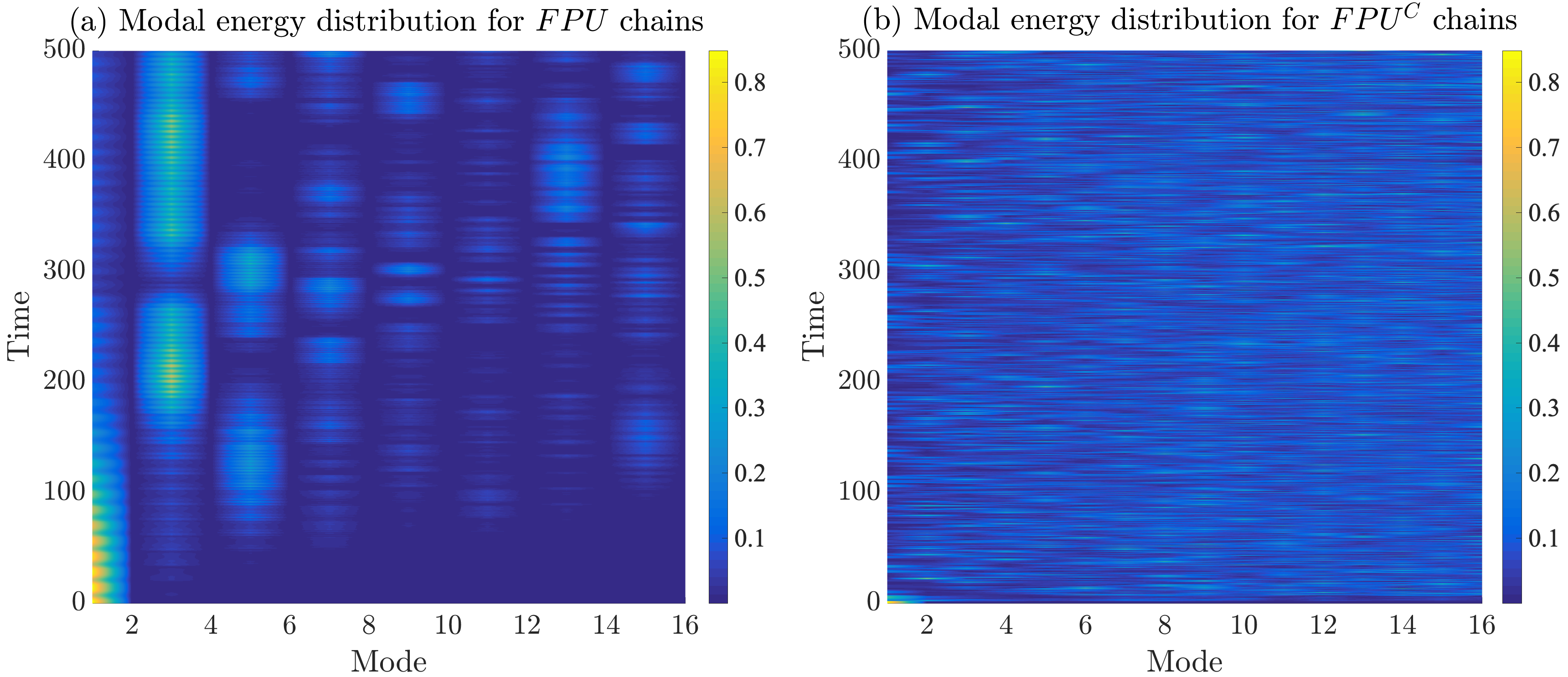} 
    \caption{Temporal evolution of the modal energy distribution with $N=16$: (a) for $FPU$ chains and (b) $FPU^C$ chains. Clearly, the energy gets transferred to the higher modes from the lowest modes in the $FPU^C$ chains much faster than the $FPU$ chains. Further, in $FPU^C$ chains all modes get excited unlike in the $FPU$ chains, where the even modes do not carry any energy.}
    \label{fig:FPUmodal1_500}
\end{figure}

Unlike in a harmonic chain, in both $FPU$ and $FPU^C$ chains, the normal modes interact and transfer energy with each other. In order to understand the energy transfer between the different modes, we take the two chains with $N=16$, impart all energy in the first mode (60 units), and continuously monitor the energy transfer between the different modes. Note that the chains are simulated without any thermostat. A quick redistribution of energy from the first mode to the higher modes indicates a strong interaction between the modes, which effectively means a shorter mean free path of the low-frequency high-wavelength modes. The shorter mean free path in turn results not only in smaller $\kappa$, but also helps a chain to obey the Fourier's laws. The temporal evolution of the modal energy as a fraction of the total initial energy is shown in figure (\ref{fig:FPUmodal1_500}) for both the chains. It is evident that the lowest modes of energy travel relatively unimpeded in the $FPU$ chain when compared with the $FPU^C$ chain. Further, the energy is quickly redistributed to all the modes of the $FPU^C$ chain, which suggests that the mean free path in the Fourier space is smaller for them. Such is not observed in $FPU$ the chain. The rate of modal energy redistribution is directly correlated to the number of collisions occurring in the $FPU^C$ chain. All of these contribute towards explaining the reduced $\kappa$ in the $FPU^C$ chains, and why it follows the Fourier's law. Interestingly, unlike in the $FPU$ chain, where only the odd modes are excited and energy equipartition does not hold true, in the $FPU^C$ chain, all modes seem to get equally excited with energy equipartition also holding true. 

\begin{figure}
    \centering
    \includegraphics[width=1.0\linewidth]{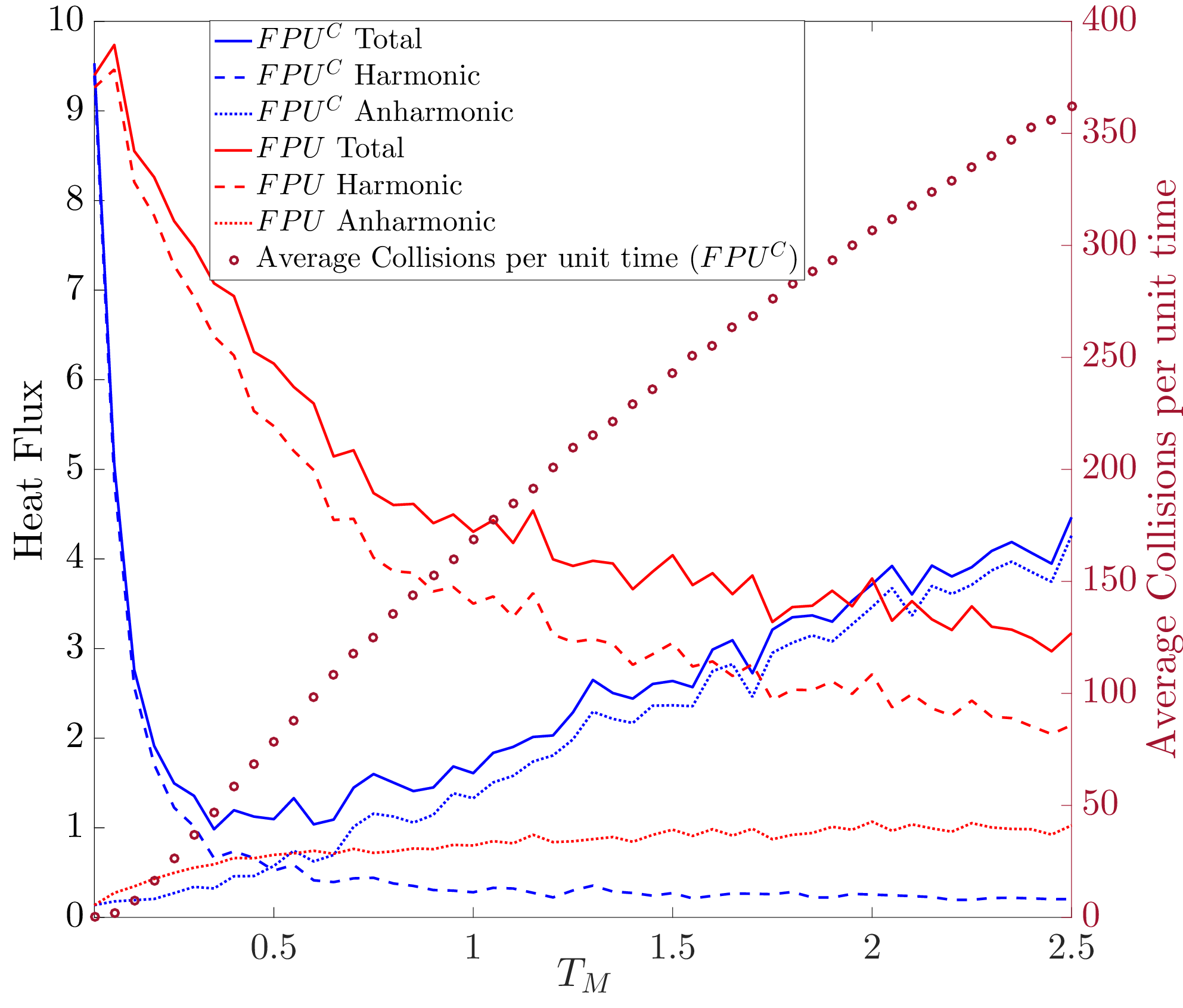}
    \caption{(Left axis)The split of total heat flux in terms of the heat fluxes due to harmonic and anharmonic forces. (Right axis) Average collisions occurring per unit time in $FPU^C$ chains. As the temperature, $T_M$ increases, unsurprisingly the harmonic contributions decrease. The anharmonic contributions on the other hand increase. The rise in anharmonic contributions may be attributed to the increase in collisions.}
    \label{fig:split_conductivity}
\end{figure}

Another interesting feature of the $FPU^C$ chains is that it has a larger $\kappa$ than the $FPU$ when $T_M \geq 2$ (see figure (\ref{fig:conductivity2}), which is difficult to explain using the arguments laid above. The origin of this feature lies in the increased contribution of the anharmonic forces towards the total heat flux at larger $T_M$. To show this we split the equation (\ref{eq:hflux}) as:
\begin{equation}
    \langle j_{i-1,i} \rangle = \langle v_if_{i-1,i}^H \rangle + \langle v_if_{i-1,i}^A \rangle,
    \label{eq:hfluxsplit}
\end{equation}
where, $f_{i-1,i}^H = -\frac{\partial V_H(\Delta x_{i-1,i})}{\partial x_i}$ and $f_{i-1,i}^A = -\frac{\partial V_A(\Delta x_{i-1,i})}{\partial x_i}$ denote the harmonic and anharmonic forces, respectively. Figure (\ref{fig:split_conductivity}) plots the variation of the total, harmonic, and anharmonic heat fluxes for both $FPU$ and $FPU^C$ chains when $T_M$ is varied from 0 to 2.5 in steps of 0.05. As $T_M$ increases, the vibration of particles around their mean position increases, and hence, the anharmonic effects start to increase as well. The increase in anharmonic contributions is much more significant in the $FPU^C$ chains than in the $FPU$ chains, predominantly because of the increased collisions.

To summarise, in this manuscript, we propose a slight modification in the traditional $FPU$ chains by incorporating a soft-sphere type collision potential in the total Hamiltonian. The resulting chain has completely different thermal transport characteristics -- the boundary jumps typically observed in the $FPU$ chains are significantly smaller, the Fourier's law is obeyed and normal thermal transport characteristics are seen. At low temperatures, the thermal conductivity of the $FPU^C$ chains is significantly smaller than that in the FPU chains owing to the quicker redistribution of the energy from the lowest modes to the higher modes. The redistribution is facilitated by the collisions between the particles of the $FPU^C$ chains. At high $T_M$, where the interaction between the modes is significant, the energy redistribution between the modes is superseded by the contribution of the short-ranged but large-magnitude collision forces towards the total heat current. The $FPU^C$ chain retains property of momentum conservation of an $FPU$ chain. This chain can, therefore, be included among \cite{savin2014thermal,giardina2000finite,gendelman2000normal,lee2010momentum,giardiana2005momentum} which further proves that momentum conservation is not a sufficient condition for anomalous thermal transport in one-dimensional chains.

\acknowledgments
PKP gratefully acknowledges the support for this research provided in part by the Indian Institute of Technology Kharagpur under the grant DNI.  
\bibliography{apssamp}
\cleardoublepage

\section{SUPPLEMENTAL MATERIAL}
\subsection{Thermostatted Equations of Motion}
The temperature control has been performed using the deterministic Nos\'e-Hoover thermostat, for which the equations of motion are:
\begin{equation}
    \begin{array}{rcl}
         \dot{x_i} = \dfrac{\partial{\mathcal{H}}}{\partial{p_i}}& , & \dot{p_i}  =  -\dfrac{\partial{\mathcal{H}}}{\partial{x_i}} - \delta_{i,1}\zeta_H p_1 - \delta_{i,N}\zeta_C p_N,\\
         \dot{\zeta_H}= \dfrac{p_1^2}{T_H} - 1 &, &\dot{\zeta_C} = \dfrac{p_N^2}{T_C} - 1,\\
    \end{array}
    \label{eq:eq5}
\end{equation}
as well as the stochastic Langevin thermostat, for which the equations of motion are:
\begin{equation}
    \begin{array}{rcl}
         \dot{x_i} & = & \dfrac{\partial{\mathcal{H}}}{\partial{p_i}} \\
         \dot{p_i} & = & -\dfrac{\partial{\mathcal{H}}}{\partial{x_i}} - \delta_{i,1}\left( p_1 - \eta_H\right) - \delta_{i,N} \left( p_N - \eta_C \right).\\
             \end{array}
    \label{eq:eq5}
\end{equation}
Here, $\delta_{i,1}$ and $\delta_{i,N}$ are the switches that take a value unity when $i=1$ and $i=N$, respectively. $\eta_{H/C}$ are the random forces whose dispersion is related to the temperature through: $\sigma_{H/C}^2 = 2 T_{H/C} / \Delta t$. While the Nos\'e-Hoover equations have been solved using the classical 4th order Runge-Kutta method, the stochastic Langevin thermostat has been solved using a Velocity-Verlet type algorithm \cite{gronbech2013simple}. For both the cases, the time step is $\Delta t = 0.0005$ time units and the simulations are run for 1 billion time steps -- the first 250 million being steady-state runs where the chain reaches steady state and the last 750 million time steps being the actual result producing runs. 

\subsection{Comparison of Temperature Profile and Boundary Jumps between Nos\'e-Hoover and Langevin thermostats}
\begin{figure}
    \centering
    \includegraphics[width=1.0\linewidth]{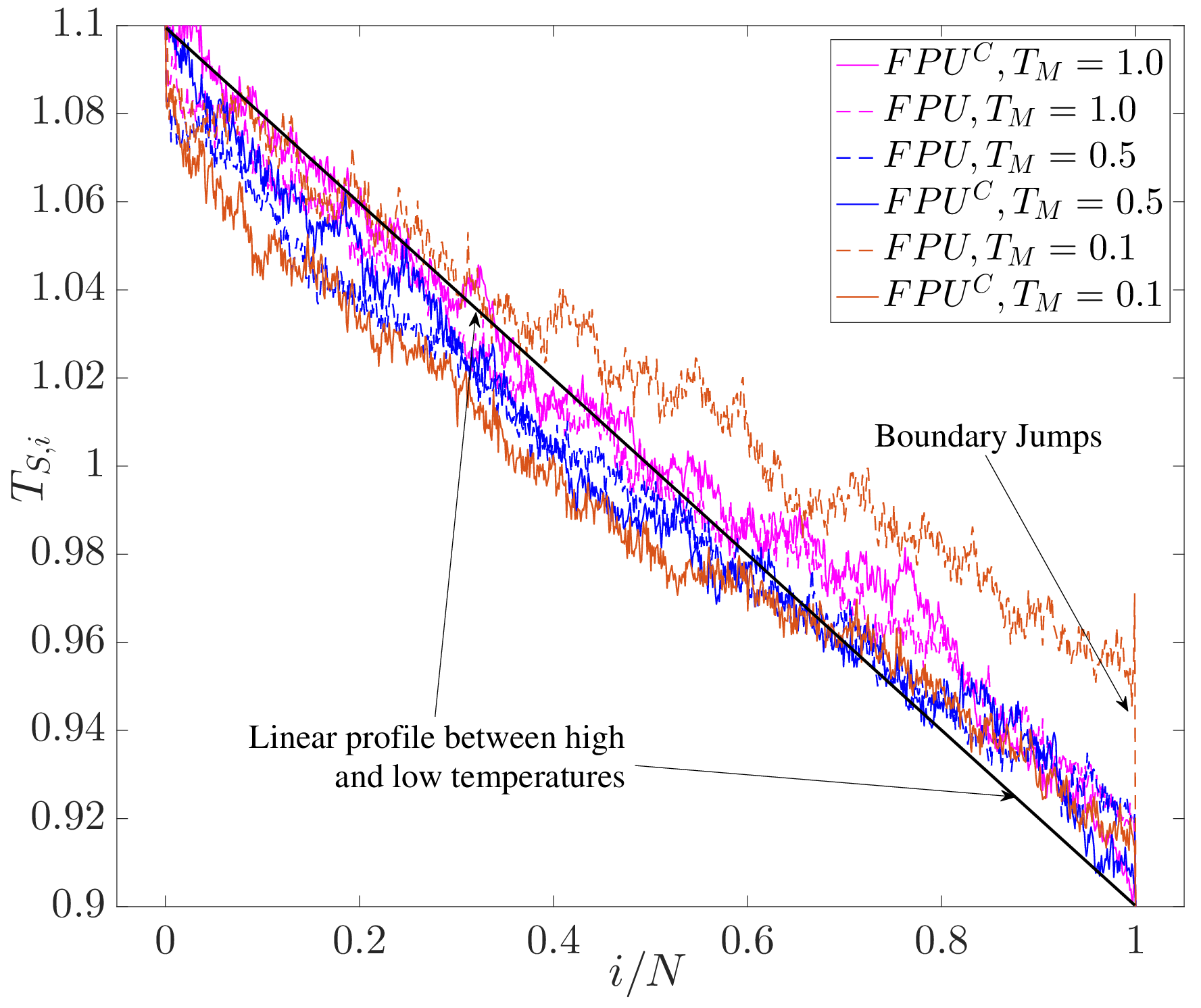}
    \caption{Temperature profile plot for the Nos\'e-Hoover thermostatted chains comprising 1024 particles. While the results are similar to the Langevin thermostatted chains (see figure (\ref{fig:TempProfile})), a noticeable difference arises for the boundary jumps, where it can be seen that the Nos\'e-Hoover thermostatted chains have significantly smaller boundary jumps.}
    \label{fig:nh_profile}
\end{figure}
The temperature profile of the chains thermostatted using the Nos\'e-Hoover thermostats has been computed using a similar method as described in the main text. First the individual particle temperatures, $T_i(t) = v_i^2(t)$, are computed which are subsequently normalized using $T_{S,i} = \langle T_i \rangle/T_M$. Figure (\ref{fig:nh_profile}) shows the normalized temperature profile for $N=1024$ particles using the Nos\'e-Hoover thermostats. When compared with figure (\ref{fig:TempProfile}), a few noticeable changes may be observed -- (i) the Nos\'e-Hoover thermostatted chains produce smaller boundary jumps than the Langevin thermostatted chains for both $FPU$ and $FPU^C$ chains, and as a result, (ii) the temperature profile of the Nos\'e-Hoover thermostatted chains are closer to linearity.

\subsection{Comparison of Thermal Conductivity between the Nos\'e-Hoover and Langevin thermostats}
\begin{figure}
    \centering
    \includegraphics[width=1.0\linewidth]{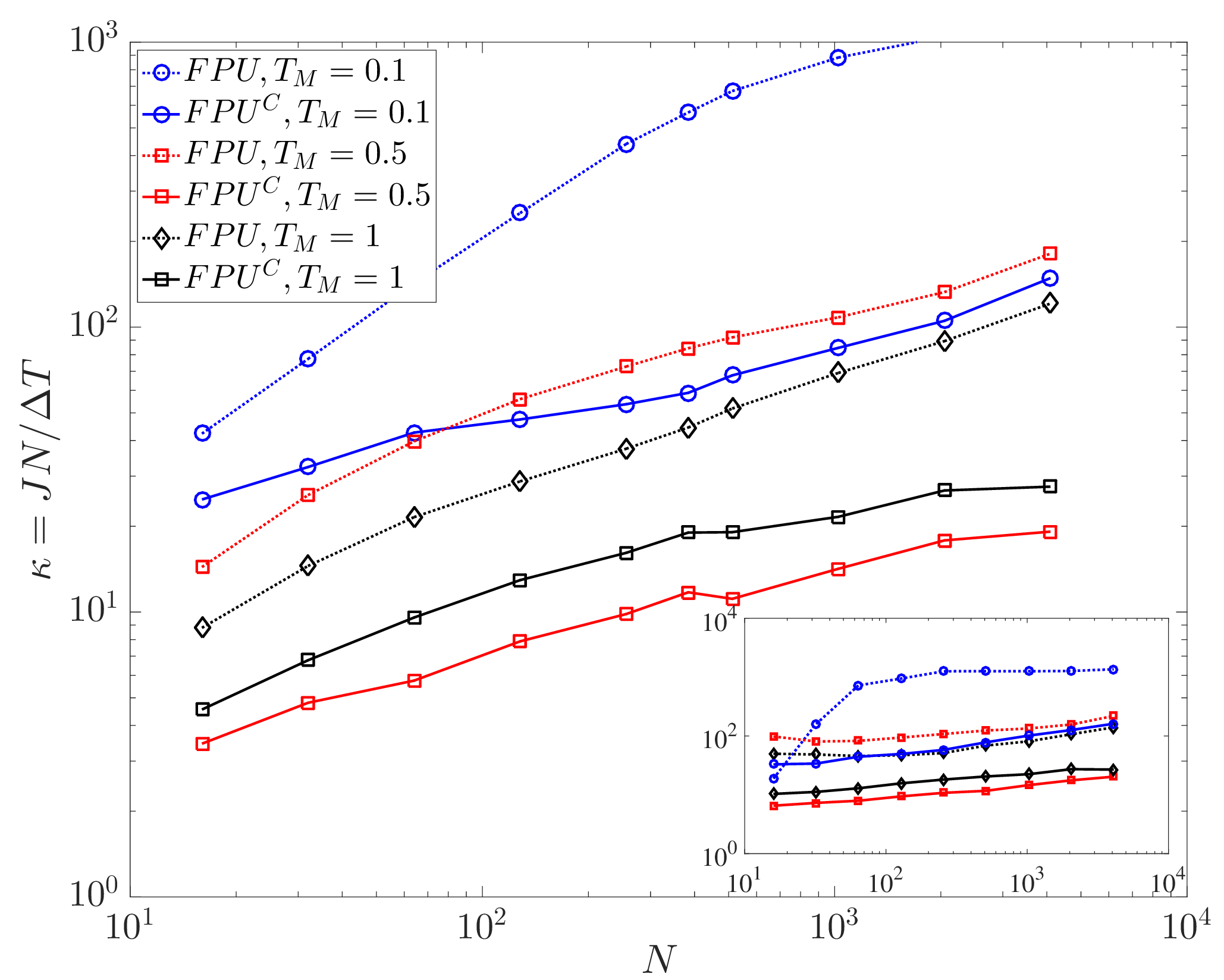}
    \caption{Thermal conductivity of $FPU$ and $FPU^C$ chains thermostatted using Nos\'e-Hoover thermostats. Two different values of $\Delta T$ have been used similar to figure (\ref{fig:conductivity2}). As the Nos\'e-Hoover chains have reduced boundary jumps, even chain with small lengths follows the overall trend.}
    \label{fig:nh_flux}
\end{figure}
The thermal conductivity of the Nos\'e-Hoover thermostatted chains has been computed using the equations (\ref{eq:hflux}), (\ref{eq:hflux2}), and (\ref{eq:hflux3}). The results are shown in figure (\ref{fig:nh_flux}). While the overall trend remains the same as observed for the Langevin thermostated chains, there are a few noticeable differences. As is the case with the Langevin thermostatted chains, we use different $\Delta T$ -- the main plot of figure (\ref{fig:nh_flux}) shows $\kappa$ computed using the imposed $\Delta T$, while the inset plot shows $\kappa$ computed using the actual $\Delta T$ between the second and the second last particles of the chains. The latter $\Delta T$ effectively ignores the boundary jumps. Note that the length of the chains used in equations (\ref{eq:hflux2}) and (\ref{eq:hflux3}) have been reduced by two for consistency. Since the boundary jumps for the Nos\'e-Hoover thermostatted chains are lower than the Langevin thermostatted chains, the difference between the two types of conductivity is small for the Nos\'e-Hoover thermostatted chains.

\end{document}